# Mechanisms of surface nanostructuring of $Al_2O_3$ and MgO by grazing incidence irradiation with swift heavy ions

M. Karlušić[1,*], R.A. Rymzhanov[2,3], J.H. O'Connell[4], L. Bröckers[5], K. Tomić Luketić[1], Z. Siketić[1], S. Fazinić[1], P. Dubček[1], M. Jakšić[1], G. Provatas[1], N. Medvedev[6,7], A.E. Volkov[2,8,9,10], M. Schleberger[5]

1 Ruđer Bošković Institute, Bijenička cesta 54, 10000 Zagreb, Croatia

2 Joint Institute for Nuclear Research, Joliot-Curie 6, 141980, Dubna, Moscow Region, Russia

3 The Institute of Nuclear Physics, Ibragimov St. 1, 050032 Almaty, Kazakhstan;

4 Nelson Mandela University, University way, Summerstrand, 6001, Port Elizabeth, South Africa

5 Fakultät für Physik and CENIDE, Universität Duisburg-Essen, D-47048 Duisburg, Germany

6 Institute of Physics, Czech Academy of Sciences, Na Slovance 2, 182 21, Prague 8, Czech Republic

7 Institute of Plasma Physics, Czech Academy of Sciences, Za Slovankou 3, 182 00 Prague 8, Czech Republic

8 National Research Centre 'Kurchatov Institute', Kurchatov Sq. 1,123182 Moscow, Russia

9 P. N. Lebedev Physical Institute of the Russian Academy of Sciences, Leninskij pr. 53,119991 Moscow, Russia

10 National University of Science and Technology MISiS, Leninskij pr., 4, 119049 Moscow, Russia

* corresponding author: marko.karlusic@irb.hr,  +385 1 456 0988

**Abstract**

We experimentally discovered that $Al_2O_3$ and MgO exhibit well-pronounced nanometric modifications on the surfaces when irradiated under grazing incidence with 23 MeV I beam, in contrast to normal incidence irradiation with the same ion beam when no damage was found. Moreover, ions in these two materials produce notably different structures: grooves surrounded with nanohillocks on MgO surfaces vs. smoother, roll-like discontinuous structures on the surfaces of $Al_2O_3$. To explain these results, detailed numerical simulations were performed. We identified that a presence of the surface inhibits recrystallization process, thereby preventing transient tracks from recovery, and thus forming observable nanopatterns. Furthermore, a difference in the viscosities in molten states in $Al_2O_3$ vs. MgO explains the differences in the created nanostructures. Our results thus provide a deeper understanding of the fundamental processes of surface nanostructuring, potentially allowing for controlled production of periodic surface nanopatterns.

1. **Introduction**

Nanometric effects caused by swift heavy ion (SHI) beams have found diverse applications such as production of track-etched membranes [1], hadron therapy [2], and testing of semiconductor devices [3]. Nanoscale damage along an SHI path is usually called an ion track and can be observed with various experimental techniques. Understanding of involved physical processes is still incomplete, and various mechanisms have been proposed for damage formation caused by these ions [4,5].

At low levels of energy deposited by an SHI, the initial excitation dissipates leaving an almost unaffected material around the ion path. Initially, it was thought that this effect could severely limit ion track research at small and medium sized accelerator facilities, but recent developments have demonstrated otherwise [6]. This paper shows that grazing incidence SHI irradiation below the bulk threshold energy of track formation can cause considerable surface damage, thus enabling this line of research at smaller accelerator facilities [7-9].

At the low-energy shoulder of the Bragg-peak, the threshold for ion tracks in $Al_2O_3$ after normal incidence of an SHI is well known: it is ~10 keV/nm in the bulk [10-12], and the threshold for nanohillocks formation observed by means of atomic force microscopy (AFM) at the surface could be extrapolated to ~10 keV/nm as well [13]. In a case of normal incidence SHI irradiation of MgO, the threshold in the bulk is above 18 keV/nm [11,14,15], while at the surface small nanohillocks have been observed already at 15 keV/nm [13].

In this work, we examine the response of these two radiation hard oxides to grazing irradiation by iodine ions with the energy losses below the track formation threshold in the bulk. By using atomic force microscopy (AFM), time-of-flight elastic recoil detection analysis (ToF-ERDA), Rutherford backscattering spectrometry in channelling mode (RBS/c) and transmission electron microscopy (TEM), we demonstrate formation of long surface tracks induced by such grazing ions. These results are supported by numerical simulations based on the TREKIS code [16,17] which illustrate different mechanisms governing the formation of SHI tracks under normal vs. grazing incidence conditions in $Al_2O_3$ vs. MgO samples.

## 2. Experimental details

Single crystal $Al_2O_3$ (0001) and MgO (100) samples were purchased from CrysTec GmbH, Germany. The samples were 5×5 mm$^2$ in size, and their thickness was 0.5 mm. The samples were epi-polished on one side, making them suitable for the AFM analysis. No sample preparation was done before the irradiation nor before the analysis except for the TEM preparation.

The samples were irradiated using 23 MeV I$^{6+}$ beam delivered from the 6 MV Tandem Van de Graaff accelerator at the Ruđer Bošković Institute (RBI) accelerator facility in Zagreb, Croatia. The crystals were irradiated in normal (5° off the surface normal to avoid channelling) and grazing (1° with respect to the surface) incidence geometries at the ToF-ERDA beamline [18]. According to the SRIM code [19], the stopping powers in MgO (d$E_e$/d$x$ = 8.53 keV/nm, d$E_n$/d$x$ = 0.42 keV/nm) and $Al_2O_3$ (d$E_e$/d$x$ = 9.1 keV/nm, d$E_n$/d$x$ = 0.45 keV/nm) are similar and the electronic component d$E_e$/d$x$ dominates the nuclear energy loss d$E_n$/d$x$.

Throughout the paper we operate with two separate quantities (and units): a) fluence [ions/cm$^2$] – the number of ions impinging per unit area, and b) surface tracks density [tracks/cm$^2$] – the number of tracks produced per unit area. These quantities are equal only for normal incidence irradiations and small fluences without significant track overlap. For irradiations out of normal incidence, due to the geometrical increase of projected exposure area, track density is equal to the fluence scaled by the sine of the irradiation angle. Therefore, a typical ion fluence of ~$3\times10^{11}$ ions/cm$^2$ used in the present work (for 1° grazing incidence angle of irradiation) results in formation of ~$5\times10^9$ tracks/cm$^2$ on the surface.

After SHI irradiations, morphologies of the surface nanostructures were examined under ambient conditions using a Dimension 3100 AFM (Veeco metrology) in the tapping mode and using NCHR cantilevers (Nanosensors, Switzerland) with cantilever resonance frequencies of around 300 kHz. Additional AFM measurements were performed using NTEGRA Prima AFM (NT-MDT spectrum instruments) in the contact mode. The obtained images were analysed using Gwyddion code [20], and from the raw data (512×512) only a parabolic background was subtracted.

The SHI irradiated samples were further analysed at the RBI accelerator facility using RBS/c and ToF-ERDA. The RBS/c was performed using 1 MeV proton beam from the 1 MV Tandetron accelerator. The beam spot size was 1 mm and the current was kept at a few nA. A silicon surface barrier detector was positioned at 160° with respect to the incoming beam direction for detection of backscattered ions [21,22]. The samples were aligned using an unirradiated part of the crystal by obtaining angular (tilt, azimuth) scan maps.

For the *in situ* ToF-ERDA, 23 MeV I$^{6+}$ beam from the Van de Graaff accelerator was used, and the measurements were conducted at the identical grazing incidence geometry, i.e. at 1° with respect to the sample surface, with the spectrometer positioned at 37.5° towards the beam direction [21]. After the collection of the data in the "list mode", offline replay and analysis were performed using the Potku code [23] following the methodology developed in Ref. [7].

Finally, selected samples were analysed with TEM. Cross sectional TEM lamellae were prepared by a standard FIB lift-out procedure using a FEI Helios 650. Prior to exposing the irradiated surface to Ga ions, a ~200 nm thick carbon layer was deposited onto the area of interest using 1 keV electrons. 30 keV Ga ions were then applied to deposit more carbon to a final thickness of 2 µm before trench milling. Final thinning was performed using 5 keV Ga and polishing at 1 keV Ga energy. TEM imaging was performed in a JEOL ARM 200F TEM with imaging Cs corrector at 200 keV energy.

## 3. Experimental results

*3.1 AFM analysis of the surface*

The inspection of the surface by an AFM reveals long ion tracks on the irradiated side of both $Al_2O_3$ and MgO samples after the grazing incidence SHI impacts. As shown in Fig. 1, the SHI irradiation yields ion tracks in the form of chains of nanohillocks on the surfaces of $Al_2O_3$ crystals. A typical height of the observed features is a few nanometers, and ion tracks have a high degree of periodicity. Larger nanodots occur due to the surface contaminations. From the applied ion fluence we estimate that the ion track production efficiency is close to one.

However, after normal incidence irradiation, $Al_2O_3$ surface remained flat even after being exposed to a high fluence of $3×10^{13}$ ions/cm². In the case of ion track formation, an increase of the surface roughness is expected [9], however no such effect was found in $Al_2O_3$ (the surface RMS roughness remained on the level of 0.1-0.2 nm).

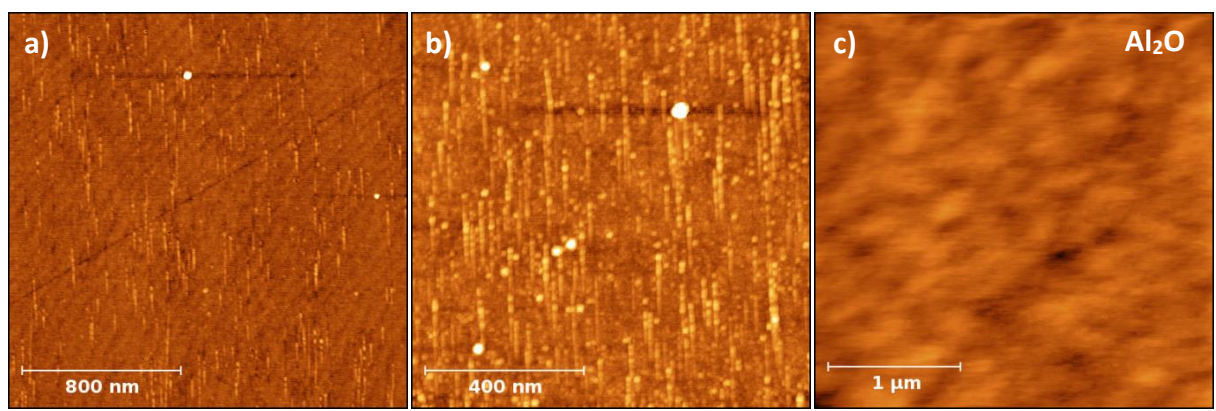

**Figure 1.** Ion tracks on $Al_2O_3$ surfaces after grazing incidence irradiation with 23 MeV I: (a) a track density of $5×10^9$ tracks/cm², and (b) $2×10^{10}$ tracks/cm². (c) Surface after normal incidence irradiation with 23 MeV $I^{6+}$ to a fluence of $3×10^{13}$ ions/cm². Height scale is (0-3) nm in all images.

In the case of MgO, similarly, ion tracks appear on the surface after grazing incidence SHI irradiation, while no effect is found after normal incidence irradiation, see Fig. 2. In this case, the morphology of the ion tracks appears to be different from the $Al_2O_3$ case. Instead of long chains of nanodots, long straight groves are observed. From the applied ion fluence, we again estimate the ion track production efficiency to be close to unity. Similarly to the case of Al$_2$O$_3$, no increase in the surface roughness is observed after normal incidence irradiations to the fluence of 3×10$^{13}$ ions/cm$^2$ (the surface RMS roughness remained on the level of 0.1-0.2 nm).

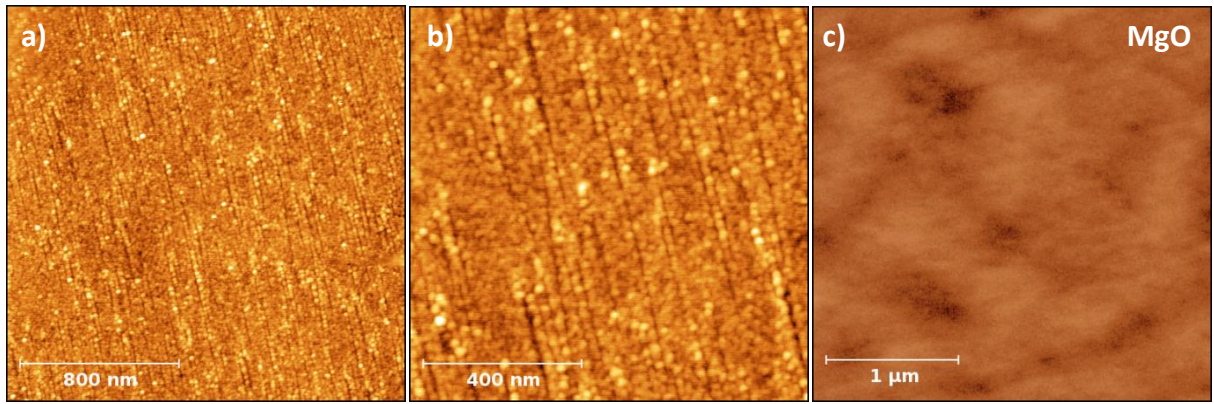

**Figure 2.** (a,b) Ion tracks on MgO surfaces after grazing incidence irradiation with 23 MeV I. A track density of 5×10$^9$ tracks/cm$^2$ is shown with different magnifications. (c) Surface after normal incidence irradiation with 23 MeV I$^{6+}$ to the fluence of 3×10$^{13}$ ions/cm$^2$. Height scale is (0-3) nm in all images.

*3.2 ToF-ERDA study of surface elemental composition*

An additional investigation of the surface track formation during the grazing SHI exposure was accomplished using *in situ* ToF-ERDA. To monitor for possible changes of the elemental composition of the surface, we performed measurements under the same irradiation conditions, i.e. the same ion type, energy and incidence angle. From the analysis of spectra shown in Fig. 3, following the methodology developed in Ref. [7] and using the Potku code [23], we have found that the stoichiometry of the surface remains unchanged in both materials even for the highest fluences of 1×10$^{12}$ tracks/cm$^2$. Obtained final elemental concentrations are the same as the initial values, i.e. in the case of Al$_2$O$_3$ were c(Al) = 38±2 at% and c(O) = 61±3 at%, while in the case of MgO concentrations were c(Mg) = 50±3 at% and c(O) = 49±3 at%. In the case of MgO, the observed gold signal comes from the gold-coated aluminium sample holder, while the iodine signal comes from the primary ion beam. In the case of Al$_2$O$_3$, a carbon coated aluminium sample holder was used.

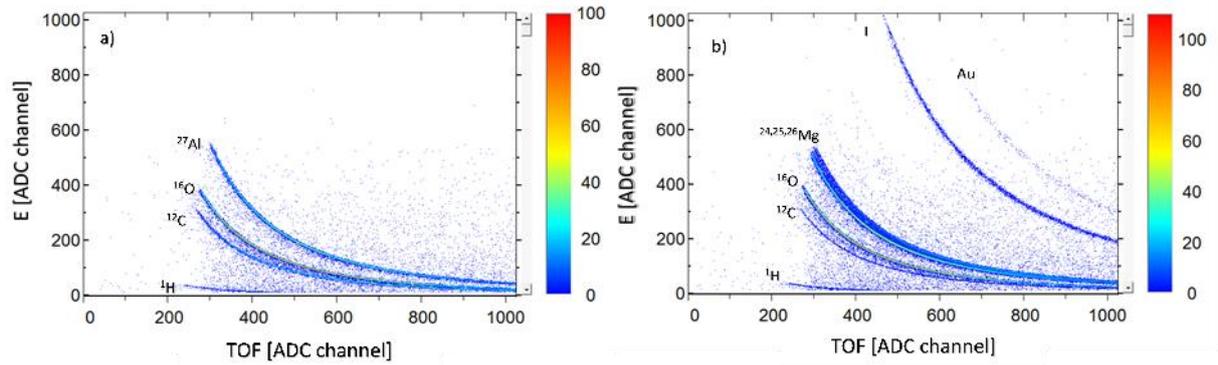

**Figure 3.** ToF-ERDA spectrum of (a) $Al_2O_3$ surface and (b) MgO surface. 2D map of the kinetic energy (E) versus the time-of-flight (TOF) of a recoil atom is shown. The colour indicates the number of events (intensity) for a given (TOF, E) pixel in a 2D map.

*3.3 RBS/c analysis of subsurface region*

RBS/c measurements were carried out to investigate possible formation of ion tracks below the surface after the normal incidence SHI irradiation. Fig. 4 shows that only a small amount of defects was formed after SHI irradiation of $Al_2O_3$ with the high fluence of $3 \times 10^{13}$ ions/cm$^2$. In a case of latent ion track formation, such a high ion fluence would have led to a significant amount of damage due to ion track overlap. The observed increase in the RBS/c yield, compared to the unirradiated sample in the case of $Al_2O_3$, is most likely due to dechanneling or backscattering from oxygen point-like defects. In the case of MgO, only a negligible amount of SHI induced defects is found. The peak observed close to the oxygen edge (around channel 520) is due to the known p-Mg resonance at 820 keV.

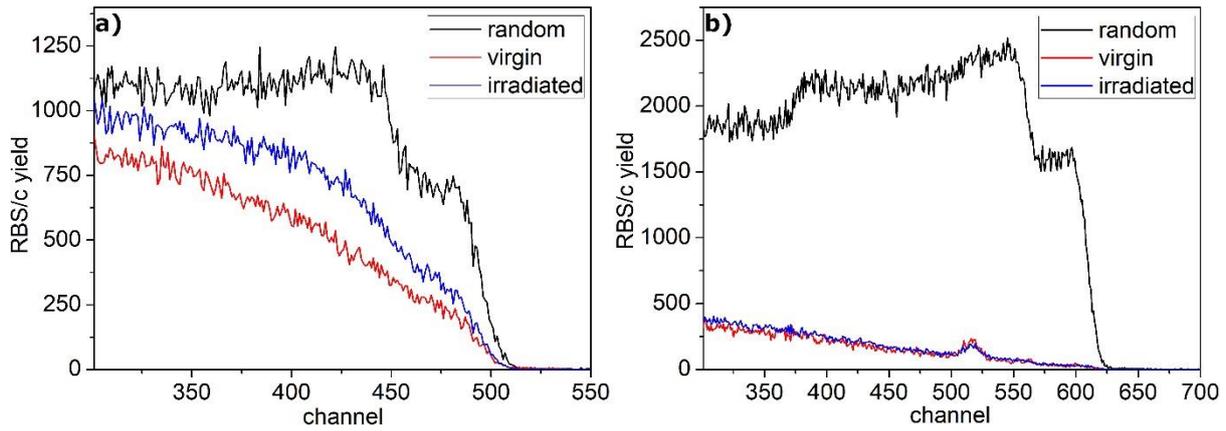

**Figure 4.** (a) RBS/c spectra from $Al_2O_3$ sample irradiated with 23 MeV I ions to the fluence of $3\times10^{13}$ ions/cm$^2$. (b) RBS/c spectra from MgO sample irradiated with 23 MeV I to the fluence of $3\times10^{13}$ ions/cm$^2$. "Random" stands for RBS spectrum, while "virgin" is RBS/c spectrum from an unirradiated sample.

*3.4 HRTEM investigation*

Finally, both MgO and $Al_2O_3$ samples irradiated with SHI at normal incidence at the low ion fluence of $2\times10^{11}$ ions/cm$^2$ were studied with transmission electron microscopy (TEM and HRTEM). The chosen fluence ensures that ion tracks (if any) would not overlap. A thin carbon foil of 20 μg/cm$^2$ (corresponding thickness is around 100 nm) was placed in front of the samples during the irradiation to ensure that the charge state of the SHI is equilibrated before the ion impact. This way, an ambiguity in the value of the SHI stopping power was eliminated. Fig. 5 shows bright field (BF) TEM micrographs of irradiated $Al_2O_3$ (a) and MgO (c) with HRTEM images of the irradiated surfaces of $Al_2O_3$ (b) and MgO (d). No evidence of ion track formation can be seen in either (a) or (c) panels. Both materials exhibit dark spot contrast due to radiation induced defects and subsequent stress fields but no aligned features indicating individual latent tracks can be discerned. The HRTEM image in the cross-section (b) shows an atomically smooth surface for $Al_2O_3$ with no visible hillocks on the surface. For MgO, it was not possible to confirm the absence or presence of hillocks since the surface was rather rough. The rough appearance was not due to lamella preparation as the crystal surface appeared visibly rough in the SEM prior to commencing lamella preparation. At present, the cause of this surface roughening is not clear but might be related to cratering due to single ion impacts. We note that this does not contradict the AFM results because those samples have been irradiated with the ion beam having 6+ charge state. In that case, due to lower charge state of the impinging SHI, the energy loss of the projectile reduces below the threshold resulting in no nanostructures at the ion impact site.

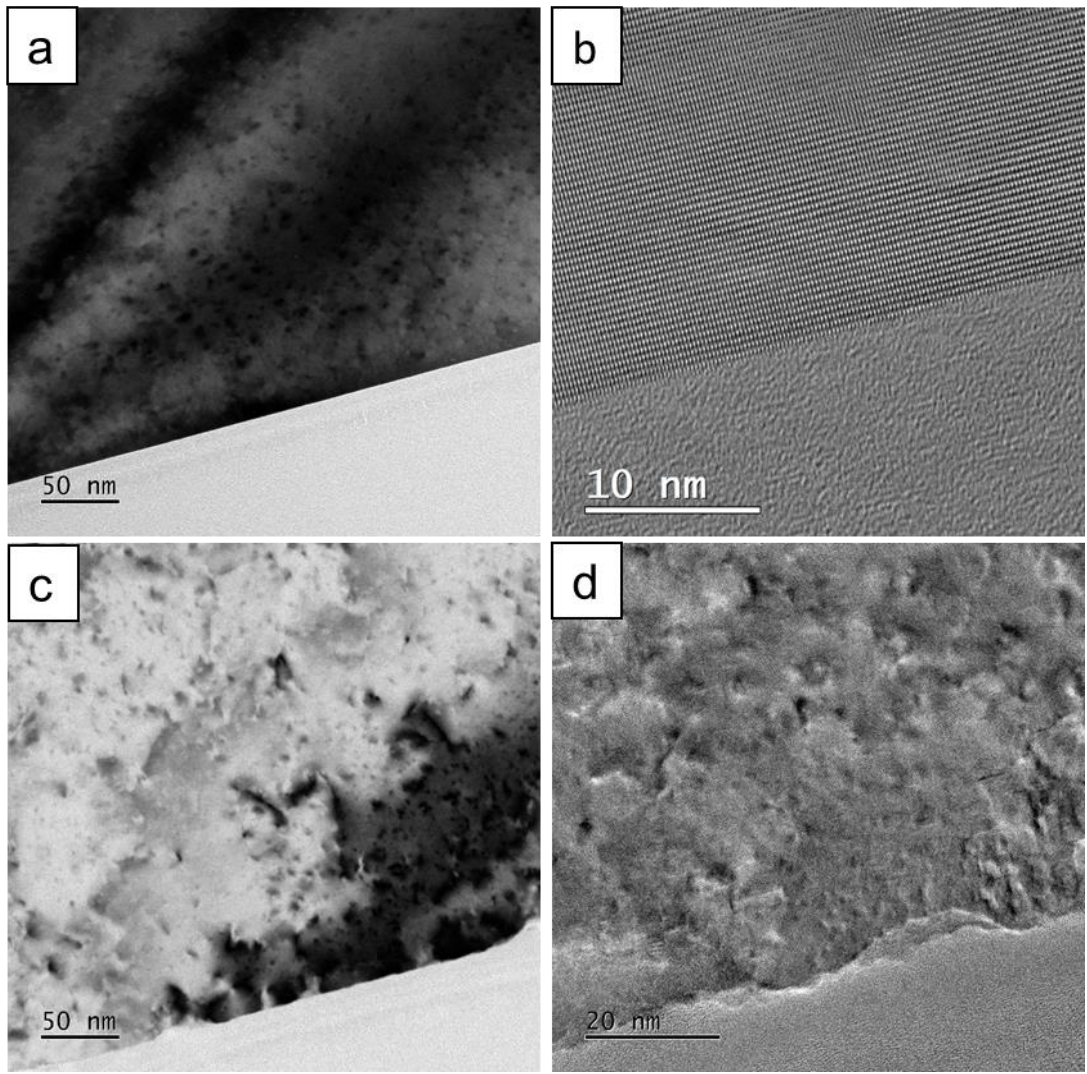

**Figure 5.** (a) TEM and (b) HRTEM micrographs of the SHI irradiated $Al_2O_3$, (c) TEM and (d) HRTEM micrographs of the SHI irradiated MgO.

## 4. Model

We applied a hybrid model combining Monte Carlo (MC) and Molecular Dynamics (MD) simulations [24] to study the coupled kinetics of excitation and relaxation of the electronic and the atomic systems of $Al_2O_3$ and MgO irradiated with SHIs under the normal and grazing incidence. First, the Monte Carlo code TREKIS (Time-Resolved Electron Kinetics in SHI-Irradiated Solids [16,17]) is used to describe SHI penetration, excitation of the ensemble of electrons, as well as energy transfer to lattice atoms via electron-lattice scattering. The cross sections applied in TREKIS are constructed within the complex dielectric function (CDF) formalism [25-27].

A calculated radial distribution of the energy transferred to the lattice forms input data for the classical molecular dynamics code LAMMPS [28] simulating subsequent lattice relaxation and structure transformations around the ion trajectory [29]. The interaction between atoms in $Al_2O_3$ and MgO is described with pair Buckingham-type potentials with parameterization taken from Ref. [30]. The supercell size used in the MD simulations of normal-incidence ion passage was 20.4×20.5×19.4 $nm^3$ (967500 atoms) in $Al_2O_3$ and 24.7×24.7×14.9 $nm^3$ (100800 atoms) with the periodic boundary conditions (PBC).

To mimic an SHI impact at a grazing angle, the ion trajectory was set parallel to the surface at 1 nm depth along *Z* axis (corresponding to ~57 nm from the starting point of an ion trajectory at the surface). For grazing-incidence simulations, box sizes were 40.1×20.5×20.5 $nm^3$ (2040000) for $Al_2O_3$ and 40.1×19.8×19.8 $nm^3$ (1751040 atoms) for MgO. To model a free surface, the box boundaries were extended by 20 nm at each side in *Z* direction, whereas *X* and *Y* borders remain under PBC.

The borders of the computational cell (last 0.5 nm) in the *X*- and *Y*-directions (only *Y*-direction for a grazing impact) were cooled to 300 K with the Berendsen thermostat [31]. The NVE (microcanonical) ensemble with 1 fs constant time-step was used in MD. The track evolution was traced up to 100-150 ps after the projectile passage. By this time, the temperature of the supercell would have decreased to below 400 K. Visualizations were made with the help of the OVITO software [32].

For an analysis of material properties, the viscosities of liquid MgO and $Al_2O_3$ at temperatures of 100 K above the melting points were calculated using molecular dynamics and the Green-Kubo formalism which relates the stress/pressure tensor to the viscosity in terms of the autocorrelation functions [33,34]. Thus, an effect of the surface tension on surface damage by an SHI was estimated. The surface tension coefficient was determined as the difference of the normal and the parallel to surface components of pressures ($\gamma = \frac{1}{2}(\sigma_{xx} + \sigma_{yy} - 2\sigma_{zz})$, where $\sigma_{ab}$ is the surface stress tensor [35]) of a liquid layer with two free surfaces [35]. For these estimations, the system was first equilibrated at a desired temperature within NPT (isothermal-isobaric) ensemble for 50 ps. Then, for evaluations of the viscosities in liquid states, the system was molten and kept at the temperature of 100 K above the melting point. The measured data were accumulated during 2 ns simulation with NVT (canonical ensemble) thermostat. Final values were obtained using double moving average of the raw data.

## 5. Simulation results

*5.1 Normal incidence of SHI impact: tracks in the bulk*

Figure 6 demonstrates MD snapshots at 100 ps after a passage of 23 MeV I ion (with an equilibrium charge state) in a bulk $Al_2O_3$ and MgO. This ion produces only a little amount of damage in both targets. A track in $Al_2O_3$ is discontinuous, has a diameter less than 1 nm and consists of small damaged crystalline regions. The passage of an ion in MgO induces formation of a number of point defects. The kinetics of tracks formation is similar to that discussed in [15] where recrystallization of an area initially molten due to an ion passage was observed in both $Al_2O_3$ and MgO. It should be noted that within the present approach the calculated threshold of track formation in the bulk $Al_2O_3$ (~6-7 keV/nm [24]) is lower than the above-mentioned experimental value of ~10 keV/nm.

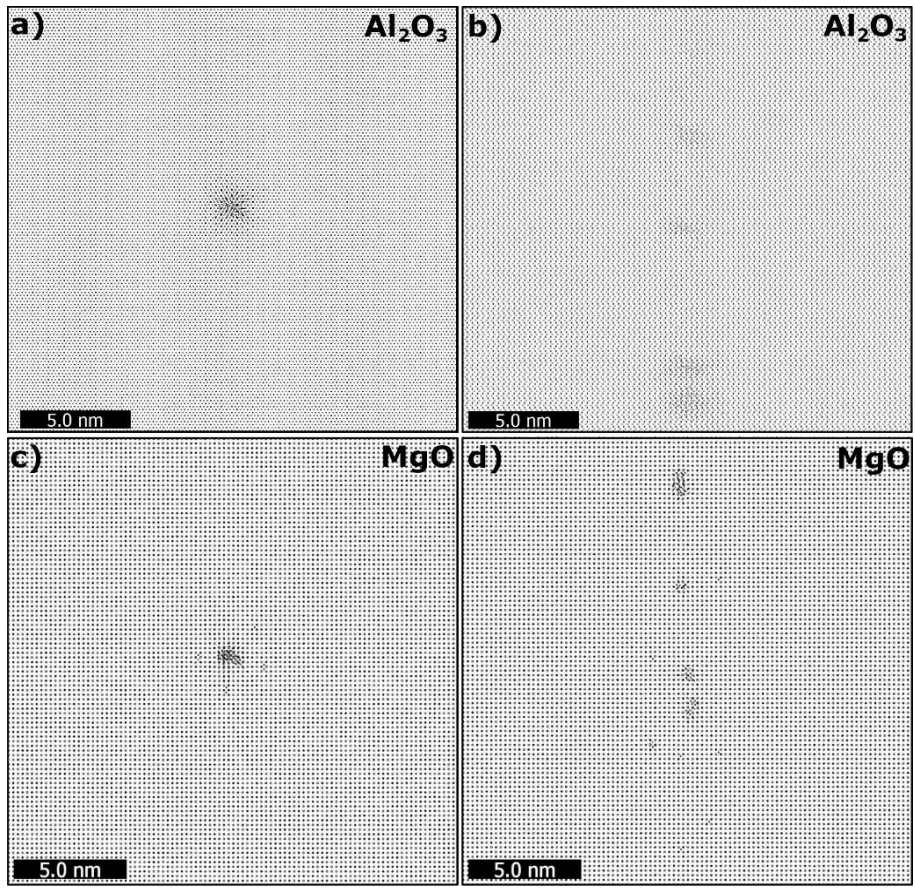

**Figure 6.** MD simulation of 23 MeV I ion track: (a) projection along the ion trajectory and (b) cross-section perpendicular to ion path in bulk $Al_2O_3$; (c) and (d) similar views in bulk MgO.

*5.2 Surface effects after normal incidence SHI irradiation*

In order to understand effects of the surface to track formation after a normal incidence ion impact, an SHI passage in a subsurface region was modelled (see Fig. 7). It is clearly seen that the presence of a surface strongly affects the ion track kinetics up to the depth of ~8 nm, forming a conically shaped track. The track structure at a larger depth is similar to that in the bulk (cf. Fig. 6b). As shown in more detail in Figure 8, surface effects increase the size of the initially damaged region in $Al_2O_3$ at times of ~3 ps, despite the fact that the initially molten zone is almost cylindrical (at ~1 ps). At longer times (~6-10 ps) when the material starts to recover, the presence of the surface considerably hinders the recrystallization process, but the efficiency of this effect decreases with depth. Figure 7a also demonstrates formation of a surface hillock of ~1.4 nm in height. In contrast, no surface hillocks were observed in the experiment (Fig. 1c). This difference is to be expected since the threshold for the track formation is underestimated in our model, as mentioned above.

The effective equilibrium charge of 23 MeV I ion according to Barkas formula [16,36] of $Z_{eff}$ = +8.47 was used in simulations presented in Fig.7a, whereas the charge state of the ion beam delivered by the accelerator (in the experiments without carbon foils) is $Z$=+6. We performed simulations of the ion passage with an artificially fixed charge state $Z$=+6 as the limiting case, which reduces the ion energy deposition with respect to the equilibrium charge by a factor of two. This simulation shows only minor damage in $Al_2O_3$ at depths up to 1.5 nm (see Fig. 7b) and no hillock appeared on the surface, indicating that the reduction of the energy deposition brings the ion into a subthreshold regime. A similar situation can be observed in MgO (Fig. 7c,d), where the equilibrium charge state produces a small crystalline hillock, while the impact of a fixed charge state $Z$=+6 ion produces almost no effect.

The kinetics of track recrystallization is presented in Fig. 8 and Fig. 9 for $Al_2O_3$ and MgO, respectively. In the case of $Al_2O_3$, the surface affects the recrystallization during formation of a hillock and a subsurface ion track, whereas the bulk part of the track is not influenced much. For comparison, the kinetics for MgO is shown in Fig. 9. It also shows that the subsurface region recrystallizes slower than the bulk part. Even at times of 20-30 ps, the nanohillock is still amorphous, but finally both the track and the nanohillock are crystalline. During the final recrystallisation phase, the size of the nanohillock does not change any more.

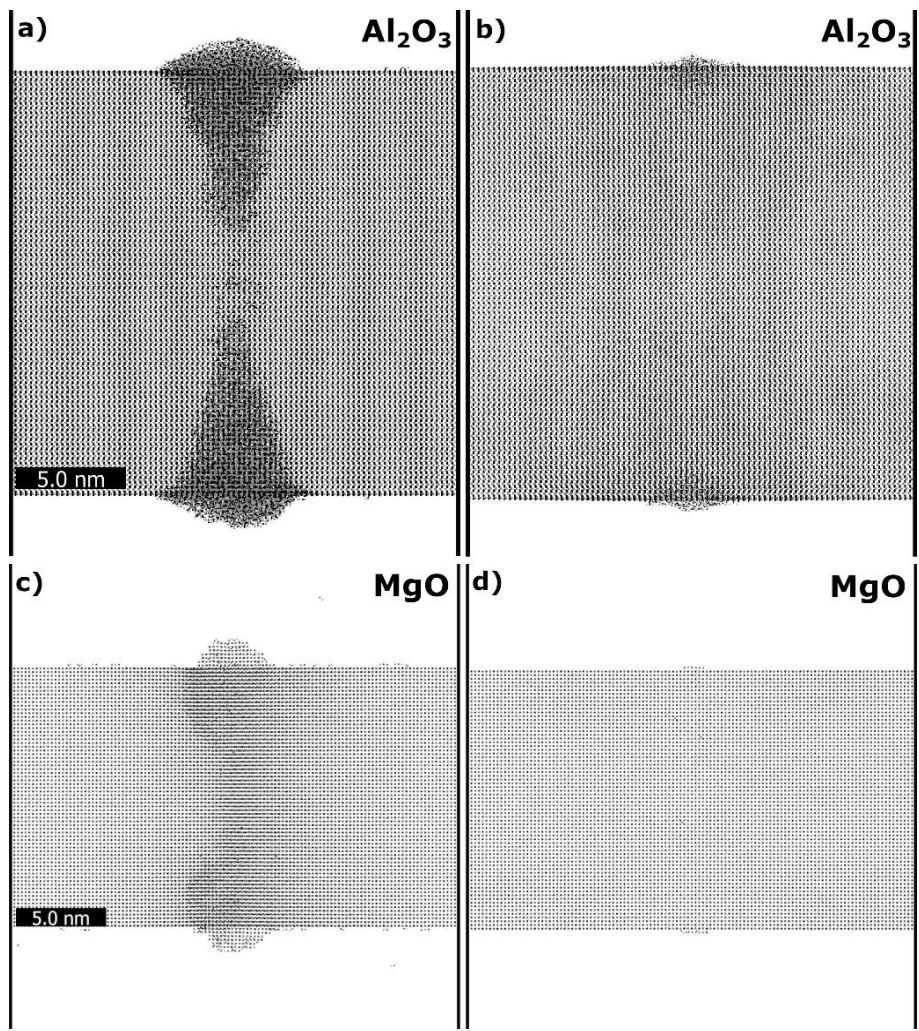

**Figure 7.** Projection along Y-axis of the MD supercells at 100 ps after the passage of 23 MeV I ion with (a) equilibrium charge state $Z_{eff}$ = +8.47 and (b) fixed charge state $Z_{eff}$ = +6 in $Al_2O_3$; (c) equilibrium charge state $Z_{eff}$ = +8.47 and (d) fixed charge state $Z_{eff}$ = +6 in MgO.

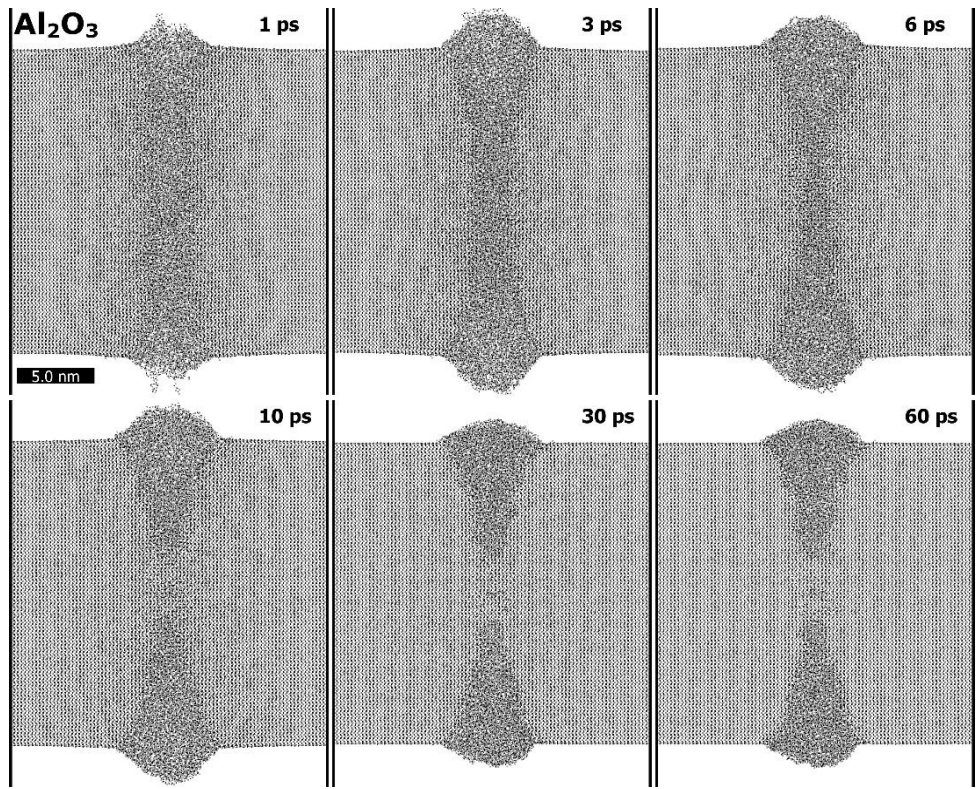

**Figure 8.** MD snapshots of a 2 nm slice of the supercell of $Al_2O_3$ irradiated with 23 MeV I (equilibrium projectile charge) at different times after the ion passage.

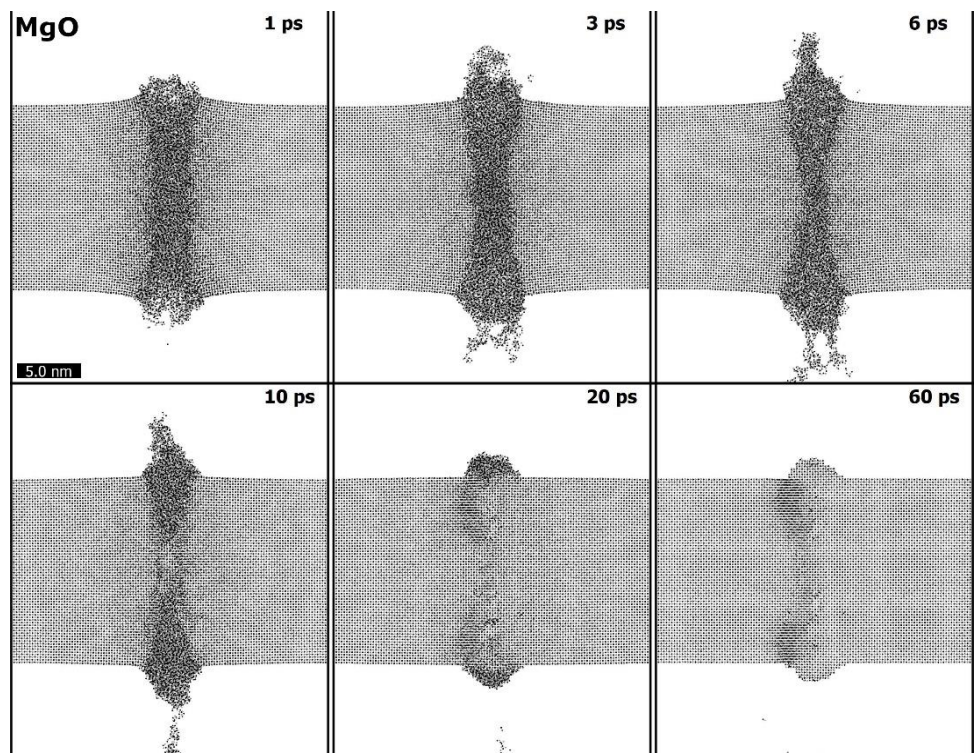

**Figure 9.** MD snapshots of a 2 nm slice of the supercell of MgO irradiated with 23 MeV I (equilibrium projectile charge) at different times after the ion passage.

*5.3 Grazing SHI incidence*

Figure 10 shows results of modelling of 23 MeV I ion passage parallel to the surface at 1 nm depth in $Al_2O_3$ and MgO. Setting the initial conditions for MD simulations, a part of the track modelled with TREKIS appearing above the surface was excluded from the subsequent calculations, effectively mimicking emission of electrons from the surface. Since the ion trajectory is much longer than the charge equilibration length, we used the equilibrium charge according to the Barkas formula.

The simulation demonstrates formation of an extended surface structure in $Al_2O_3$ (Fig. 10a), which consists of small extended hillocks with a height of about 1.5-2 nm. In contrast to $Al_2O_3$, MgO shows formation of a grove-like extended structure surrounded by droplets of ejected material (Fig. 10b). Both structures are in a good agreement with the experimental observations (cf. Figs. 1 and 2 above).

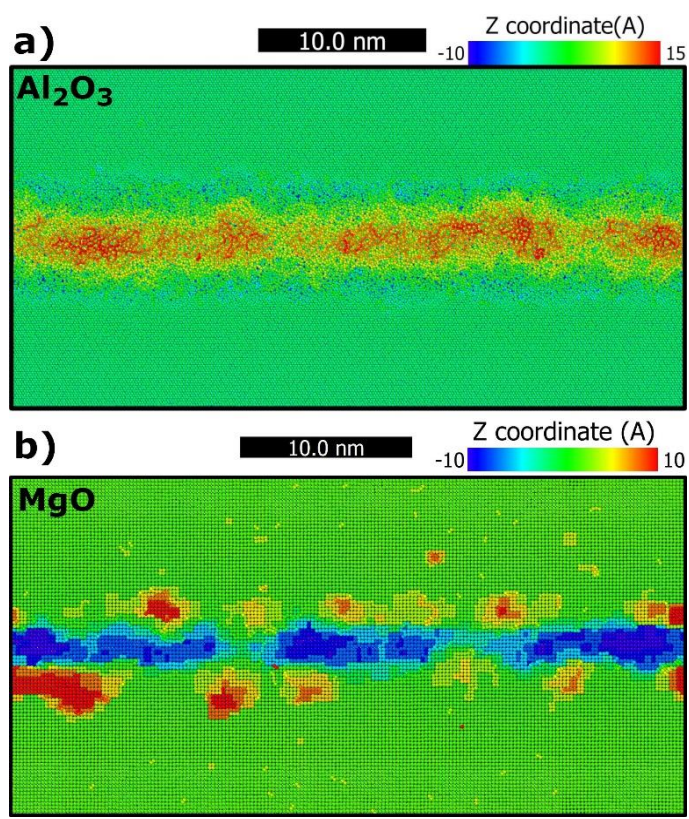

**Figure 10.** Projection of a simulation box along Z axis at 100 ps after 23 MeV I ion passage at a depth of 1 nm parallel to the surface in a) $Al_2O_3$ and b) MgO. Atoms are coloured according to their Z coordinate, while surface is at Z=0.

Fig. 11 shows the kinetics of formation of surface defects in MgO upon irradiation at grazing angles (see also corresponding animation in the Supplementary Materials). At the time of 3 ps after the passage of the ion, an expanding cylinder with a shell of molten material forms, which then bursts (at 5-10 ps) separating extruded material in opposite directions along the ion path. The process is accompanied by a significant emission of atoms and clusters from the track region at later times (~10 ps). A part of the material is deposited on both sides of the groove where it recrystallizes together with the near-surface MgO region at times > 20 ps.

The kinetics of surface structure formation in $Al_2O_3$ (Fig. 12) is relatively simple in comparison to that in MgO. A passage of an SHI causes slight protrusion of the material with minor sputtering of atoms, and this protrusion then cools down forming a roll-like structure. The complete animation can be found in the Supplementary Materials.

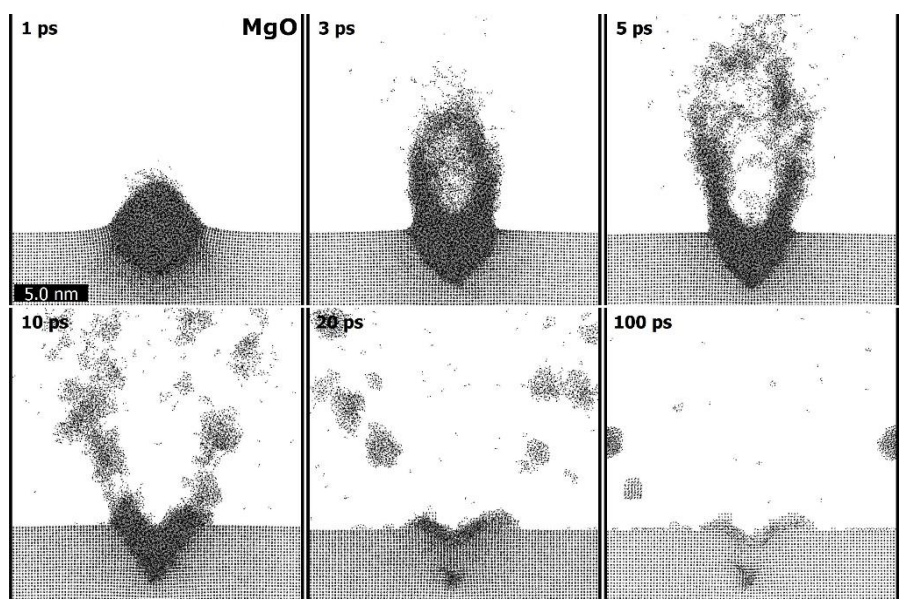

**Figure 11.** Kinetics of surface rift formation in MgO at different times after 23 MeV I ion passage at the depth of 1 nm parallel to the surface. The projection along the ion trajectory is shown.

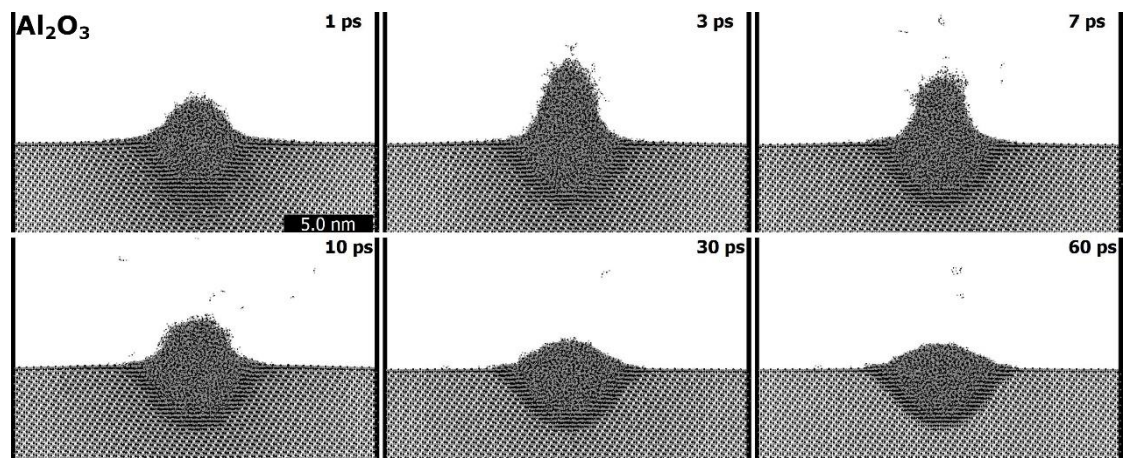

**Figure 12.** Kinetics of surface defects formation in $Al_2O_3$ at different times after 23 MeV I ion passage at the depth of 1 nm parallel to the surface. Projection along the ion trajectory is shown.

The difference in behaviour of these two materials under grazing SHI irradiation can be attributed to the mobility of atoms in the molten state. In order to interpret this effect, Table 1 shows viscosities of atoms in molten MgO and $Al_2O_3$ calculated with MD at temperatures of 100 K above the melting points (see details in Model Section). Higher mobilities of atoms in molten MgO can be noted (viscosity is several times lower in magnesia). A lower viscosity of atoms in MgO allows the material to be expelled easier from the target surface in the vicinity of the SHI trajectory. A higher viscosity in $Al_2O_3$ results in minimal extrusion of the material from a track.

Another factor hindering material expulsion from a surface of a liquid is the surface tension. MD calculations at 100 K above the melting points show that higher surface tension in $Al_2O_3$ (Table 1) limits extrusion of a liquid material from the surface, transiently forming a roll-like hillock chain. In contrast, a lower surface tension in MgO allows material extrusion at early times and, since a part of the material is ejected, a groove is left behind.

| Material | MD calc. viscosity, mPa·s | Surface tension, mN/m |
|---|---|---|
| MgO | 1.44 | 325.0 |
| $Al_2O_3$ | 9.72 | 594.8 |

**Table 1.** Viscosity and surface tension of molten $Al_2O_3$ and MgO calculated with the use of molecular dynamics.

## 6. Discussion

Careful investigations of SHI irradiated $Al_2O_3$ demonstrated that ion track formation in $Al_2O_3$ is a two steps process starting from the threshold of 10 keV/nm [12,37] and leading to the full amorphization at high fluences [38]. In the case of MgO, less research has been done, but this material is known to be highly resistant to electronic excitations induced by an SHI impact. A very high threshold for ion track formation (above 18 keV/nm) has been established for MgO [11,14,15,39].

Grazing incidence SHI irradiation makes identification of the SHI impact sites much easier, allowing studies of unusual track morphologies and processes occurring during the SHI-surface interaction [7,40-45]. For example, an origin of chainlike morphology of surface ion tracks has been attributed to the oscillating stopping power when SHI encounters different electronic densities when passing through the crystal layers at grazing incidence angles [40]. However, this effect could also play a role in ion track formation within the bulk, especially close to the threshold, and under the near-channelling irradiation conditions. An angle-dependent threshold is a clear evidence of oscillating stopping power causing ion track formation under the near-channelling conditions [9].

There are many other processes that can either lower or raise the threshold for ion track formation on the surface. First, estimations made in Ref. [46] show that a surface may localize material excitation lowering the threshold for ion track formation. Second, electrons ejected into the vacuum can carry away a portion of the SHI deposited energy, but could also promote a Coulomb explosion mechanism due to (at least temporary) charge imbalance [47]. Third, ejection of the material (electronic sputtering) can also be active at high SHI energies and lead to another channel of energy dissipation [48]. Fourth, a presence of a surface can impede recrystallization process that is effective in the bulk [7].

Generally, it has been experimentally observed that the surface ion tracks found after grazing incidence SHI irradiation have significantly lower thresholds than ion tracks in the bulk in the same material [7,9,43]. In this sense, present results agree with this observation, i.e. 8-9 keV/nm is a sufficient electronic stopping to produce well-developed ion tracks on the $Al_2O_3$ and MgO surfaces.

The main experimental result in the case of $Al_2O_3$ is an absence of ion tracks in the bulk at 9 keV/nm, and an appearance of long ion tracks at the surface after grazing incidence irradiation at the same electronic stopping power. The present study demonstrates that recrystallization plays a major role in ion track formation in $Al_2O_3$. The surface can suppress this process even at normal incidence irradiation. MD simulations clearly show the suppression of recrystallization during track formation in $Al_2O_3$ (see Fig. 8), leading to the experimentally detected ion track morphology already at 9 keV/nm.

To explain the appearance of grooves on MgO surfaces, the suppression of recrystallization alone is not sufficient. Previously, grooves have been observed in SiC, another material highly resistant to SHI irradiation [43]. Sublimation of silicon atoms has been proposed as the formation mechanism of these shallow grooves. More recently, grooves found in $CaF_2$ and mica have also been linked to the sublimation of the target [44,49], and material decomposition has been attributed to grooves formation in GaN [7]. As shown in Fig. 11, MD simulations indicate that a splash of molten MgO following the SHI passage results in a groove along the ion trajectory and a part of displaced material can be found at the track rim. Also, ToF-ERDA measurements show that the surface stoichiometry remains unchanged. Therefore, processes leading to the observed morphology of the surface ion tracks in MgO are governed primarily by the properties of the melt, and not by the recrystallization process.

**Conclusion**

In this combined experimental and theoretical study, we explored reasons why ion tracks on the materials surfaces are more easily formed than in the bulk. We showed that both in $Al_2O_3$ and MgO an effect of recrystallization makes ion tracks smaller and increases the threshold for their formation in the bulk vs. the surface. To some extent, simulations indicate that the ion charge state can influence the threshold for nanohillock formation as well.

This study identified that properties of the molten material play a significant role in track formation on the surface, especially in the case of grazing incidence SHI irradiation.

Different track morphologies in MgO vs. $Al_2O_3$ were attributed to differences in atomic mobilities. In the case of $Al_2O_3$, the suppression of recrystallization at the material surface lowers the threshold for ion track formation, but a usual ion track morphology as a sequence of nanohillocks produced along an ion trajectory was found. Compared to $Al_2O_3$, a lower viscosity and lower surface tension in molten MgO leads to material loss, as clearly demonstrated by simulations, which finally results in the grove-like morphology of ion tracks found experimentally.

**Acknowledgements**

This work was supported by the Croatian Science Foundation (HRZZ pr. no. 2786 and 8127). The authors acknowledge financial support from the European Regional Development Fund for the 'Center of Excellence for Advanced Materials and Sensing Devices' (Grant No. KK.01.1.1.01.0001)". The authors acknowledge CERIC-ERIC Consortium for the access to experimental facilities and financial support. This work is partially supported by the IAEA research contract No. 23009. L.B. and M.S acknowledge support from the DFG within the SFB1242 "Non-Equilibrium Dynamics of Condensed Matter in the Time Domain" (project number 278162697). Partial financial support from the Czech Ministry of Education, Youth and Sports, Czech Republic grants LTT17015, LM2015083 is acknowledged by N. Medvedev. The work was supported in parts by the Ministry of Science and High Education of the Russian Federation in the frameworks of FAIR-Russia Research Center (FRRC) and No. 16 APPA (GSI) as well as Increase Competitiveness Program of NUST «MISiS», Moscow, Russia under Grant K3-2018-041. The work of A.E. Volkov was supported by NRC Kurchatov Institute (n.1603). R.A. Rymzhanov acknowledges partial financial support from AYSS JINR grant No. 20-502-06. This work has been carried out using computing resources of the Federal collective usage center Complex for Simulation and Data Processing for Mega-science Facilities at NRC "Kurchatov Institute", http://ckp.nrcki.ru/, as well as computing resources of GSI Helmholtzzentrum https://www.gsi.de/en/work/research/it/hpc) and the HybriLIT heterogeneous computing platform (LIT, JINR, http://hlit.jinr.ru).